\documentclass{aa}
\usepackage{graphics}
\newcommand{\cmtwo}{cm$^{-2}$}  
\newcommand{\cmthree}{cm$^{-3}$}

\newcommand{\kms}{km\,s$^{-1}$}       
\newcommand{\vlsr}{$v_{\rm LSR}$}        

\newcommand{\dv}{$\Delta v$}

\newcommand{\um}{$\mu$m}                                 
\newcommand{\molh}{H$_{2}$}                              

\newcommand{\about}{$\sim$}                       
\newcommand{\powten}[1]{10$^{#1}$}
\newcommand{\ro}{$\rho \, {\rm Oph}$}

\newcommand{\roa}{$\rho \, {\rm Oph \, A}$}

\newcommand{\amin}{$^{\prime}$}                   
\newcommand{\asec}{$^{\prime \prime}$}
\newcommand{\adeg}{$^{\circ}$}

\newcommand{\radot}[4]{\mbox{#1$^{\rm h}$#2$^{\rm m}$#3$\stackrel{\rm s}
{_{\bf\cdot}}$#4}}  

\newcommand{\decdot}[4]{\mbox{#1$^{\circ}$ #2$^{\prime}$ #3$\stackrel {\prime 
\prime}{_{\bf \cdot}}$#4}}

\newcommand{\amindot}[2]{\mbox{#1$\stackrel {\prime}{_{\bf \cdot}}$#2}}
\newcommand{\asecdot}[2]{\mbox{#1$\stackrel {\prime \prime}{_{\bf \cdot}}$#2}}

\begin{document}

   \title{First detection of NH$_3$ ($1_0 \rightarrow 0_0$) from a low mass cloud core\thanks{Based 
  on observations with Odin, a Swedish-led satellite project funded jointly by 
  the Swedish National Space Board (SNSB), the Canadian Space Agency (CSA), 
  the National Technology Agency of Finland (Tekes)  and Centre National
  d'Etude Spatiale (CNES). The Swedish Space Corporation has been the industrial 
  prime contractor.}
  \thanks{and on observations collected with the Swedish ESO
  Submillimeter Telescope, SEST, in La Silla, Chile.}
  }

   \subtitle{On the low ammonia abundance of the $\rho$\,Oph\,A core}

   \author{     R.\,Liseau\inst{1}    	  \and
		B.\,Larsson\inst{1}       \and
		A.\,Brandeker\inst{1}     \and
            	P.\,Bergman\inst{2}    	  \and
		P.\,Bernath\inst{3}       \and
            	J.H.\,Black\inst{2}    	  \and 
		R.\,Booth\inst{2}         \and
	    	V.\,Buat\inst{4}    	  \and
		C.\,Curry\inst{3}    	  \and
		P.\,Encrenaz\inst{5}      \and 		 
		E.\,Falgarone\inst{6}     \and
                P.\,Feldman\inst{7}       \and 
		M.\,Fich\inst{3}    	  \and
                H.\,Flor\'{e}n\inst{1}    \and
		U.\,Frisk\inst{8}  	  \and  
		M.\,Gerin\inst{6}    	  \and
		E.\,Gregersen\inst{9}     \and 
		J.\,Harju\inst{10}    	  \and		
                T.\,Hasegawa\inst{11}     \and
		\AA.\,Hjalmarson\inst{2}  \and
                L.\,Johansson\inst{2}     \and
		S.\,Kwok\inst{11}    	  \and
                A.\,Lecacheux\inst{12}    \and
		T.\,Liljestr\"om\inst{13} \and
                K.\,Mattila\inst{10}      \and
		G.\,Mitchell\inst{14}     \and
		L.\,Nordh\inst{15}        \and
		M.\,Olberg\inst{2}    	  \and
		G.\,Olofsson\inst{1}      \and
		L.\,Pagani\inst{5}	  \and
		R.\,Plume\inst{11}	  \and
		I.\,Ristorcelli\inst{16}  \and 
                Aa.\,Sandqvist\inst{1}    \and
                F.v.\,Sch\'eele\inst{8}   \and
		G.\,Serra\inst{16}        \and
		N.\,Tothill\inst{14}      \and
                K.\,Volk\inst{11}         \and
		C.\,Wilson\inst{9} 	  
	}

   \offprints{R. Liseau}

   \institute{ Stockholm Observatory, SCFAB, Roslagstullsbacken 21, SE-106 91 Stockholm, Sweden \\
   	\email{rene.liseau@astro.su.se}
    \and  
         Onsala Space Observatory, SE-439 92, Onsala, Sweden         
    \and         
         Department of Physics, University of Waterloo, Waterloo, ON N2L 3G1, Canada
    \and      
         Laboratoire d'Astronomie Spatiale, BP 8, 13376 Marseille CEDEX 12, France
    \and       
    	LERMA \& FRE 2460 du CNRS, Observatoire de Paris, 61, Av. de l'Observatoire, 75014 Paris, France
    \and
    	LERMA \& FRE 2460 du CNRS, Ecole Normale Sup\'erieure, 24 rue Lhomond, 75005 Paris, France
    \and
	Herzberg Institute of Astrophysics, 5071 West Saanich Road, Victoria, BC, V9E 2E7, Canada
    \and       
    	Swedish Space Corporation, P O Box 4207, SE-171 04 Solna, Sweden   
    \and    	
    	Department of Physics and Astronomy, McMaster University, Hamilton, ON, L8S 4M1, Canada
    \and    	
    	Observatory, P.O. Box 14, University of Helsinki, 00014 Helsinki, Finland
    \and    	
    	Department of Physics and Astronomy, University of Calgary, Calgary, ABT 2N 1N4, Canada
    \and
	LESIA, Observatoire de Paris, Section de Meudon, 5, Place Jules Janssen, 92195 MEUDON CEDEX, France
    \and   
    	Mets\"ahovi Radio Observatory, Helsinki University of Technology, Otakaari 5A, FIN-02150 Espoo, Finland    
    \and      
    	Department of Astronomy and Physics, Saint Mary's University, Halifax, NS, B3H 3C3, Canada
    \and      
        Swedish National Space Board, Box 4006, SE-171 04 Solna, Sweden
    \and    
    	CESR, 9 Avenue du Colonel Roche, B.P. 4346, F-31029 Toulouse, France
    }

\date{Received date: \hspace{5cm}Accepted date:}

   \abstract{Odin has successfully observed the molecular core $\rho$\,Oph A in the 572.5\,GHz rotational ground state line of
		ammonia, NH$_3$ ($J_K = 1_0 \rightarrow 0_0$). The interpretation of this result makes use of complementary
		molecular line data obtained from the ground (C$^{17}$O and CH$_3$OH) as part of the Odin preparatory work. 
		Comparison of these observations with theoretical model calculations of line excitation and transfer
		yields a quite ordinary abundance of methanol,
		$X$(CH$_3$OH)$= 3 \times 10^{-9}$. Unless NH$_3$ is not entirely segregated from 
		C$^{17}$O and CH$_3$OH, ammonia is found to be significantly underabundant with respect to typical dense core values, 
		viz. $X$(NH$_3) = 8 \times 10^{-10}$.
               }
   \maketitle

%

\section{Introduction}

The ammonia molecule (NH$_3$) has a complex energy level structure, which makes it a useful tool to 
probe regions of very different excitation conditions in a given source (see \cite{hoandtownes83}, who also provide 
an energy level diagram), and molecular clouds have routinely been observed from the ground in the inversion lines
at about 1.3\,cm, with critical densities of about \powten{3}\,\cmthree. On the other hand, 
the rotational lines have much shorter lifetimes (minutes compared to months) and consequently much higher 
critical densities ($>10^7$\,\cmthree, see Table\,\ref{nh3_model}). Their wavelengths fall, however, 
into the submillimeter and far infrared regime and these lines are generally not accessible from the ground.
Using the Kuiper Airborne Observatory (KAO), the submillimeter ground state line of ammonia of wavelength 524\,\um, 
NH$_3$ ($J_{K}= 1_0 \rightarrow 0_0$), was first and {\it solely} detected 20 years ago toward Orion\,OMC1 
by \cite{keene83}. Only recently have renewed attempts been made to observe this line
with the spaceborne submillimeter telescope Odin (Frisk et al., Hjalmarson et al., Larsson et al., Nordh et al. and 
Olberg et al., this volume).

The cold and dense molecular core \roa\ is part of the \ro iuchi cloud at the distance of 160\,pc 
(\cite{loren90}), which is a region of ongoing low mass star formation. 
In this {\it Letter}, we present Odin observations of \roa\ in the NH$_3$\,572.5\,GHz rotational ground state
transition. Compared to the KAO, the highly improved sensitivity of Odin permits the clear detection of
this tenfold weaker line. These Odin observations were complemented with C$^{17}$O and CH$_3$OH data
obtained with the Swedish ESO Submillimeter Telescope (SEST) in La Silla, Chile (see Table\,\ref{sest_mol}), 
aimed at the determination of the average physical conditions
of the \roa\ core and these are discussed in Sect.\,4.1. The Odin observations and their results are presented 
in Sects.\,2 and 3. The implications are discussed, together with our conclusions, in Sect.\,4.2. 

\section{Odin observations and data reductions}
 
Odin observed \roa\ in NH$_3$ on February $13-15,\,2002$, toward RA\,=\,\radot{16}{26}{29}{78} 
and Dec\,=\,\decdot{$-24$}{23}{42}{3} (J2000) with a 2\amin\ FWHM circular beam.
The pointing was accurate to 30\asec, 
with an rms-stability during these observations 
better than 5\asec\ ($\Delta {\rm RA}$\,=\,$\pm$\asecdot{4}{7} and $\Delta {\rm Dec}$\,=\,$\pm$\asecdot{1}{1}).
The data were obtained in a sky-switching mode (Olberg et al., this volume) by observing 
a total of 4500\,s each ON-source and on blank sky, with 10\,s 
integrations per individual scan. In addition, once per orbital revolution, an OFF-position (1\adeg\ north) was observed, 
for a total integration time of 7000\,s.

\begin{figure}[t]
  \resizebox{\hsize}{!}{
  \rotatebox{00}{\includegraphics{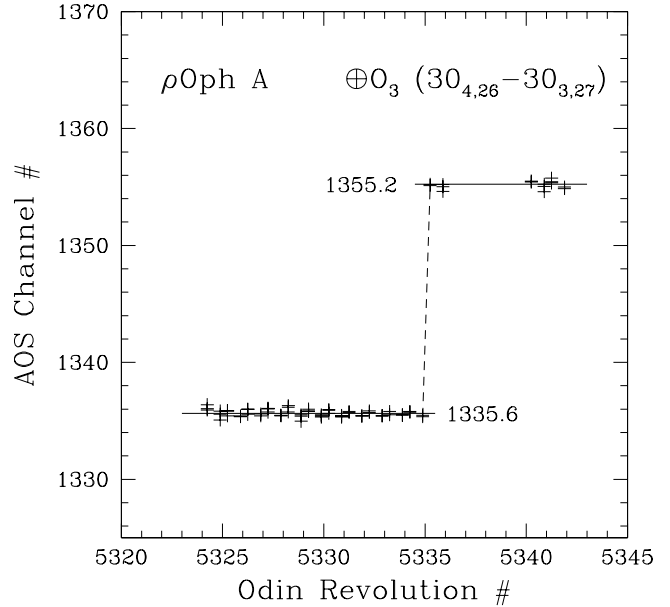}}
                        }
  \caption{Gaussian fits to the weak telluric O$_3$\,($30_{4,\,26}-30_{3,\,27}$) line
  during the Odin observations of \roa\ determined the AOS channel of the line peak 
  (channel width =\,0.32\,\kms). The adopted average channel values for the segments, during 
  which the phase lock was stable, are indicated (see the text).}
  \label{O3_phase}
\end{figure}

Shortly after launch, it was recognised that the 572\,GHz Schottky receiver was not properly phase-locked. 
However, Odin `sees' the Earth's atmosphere during its revolution and
a relatively weak telluric ozone line, O$_3$ ($J_{K^-,\,K^+}= 30_{4,\,26} \rightarrow 30_{3,\,27}$) 572877.1486\,MHz 
(\cite{pickett98}), falls sufficiently close to the ammonia line,
NH$_3$ ($J_{K}= 1_0 \rightarrow 0_0$) 572498.0678\,MHz, to allow monitoring of the receiver frequency. For large
portions of the \roa\ observations, the O$_3$ line center frequency was within 0.3 of a spectrometer channel 
(AOS-channel width = 0.32\,\kms)
and we used these fiducial channels to restore the remaining data in frequency space (cf. Fig.\,\ref{O3_phase}). 
Data collected during revolution 5336 to 5339 were not used because of too-low mixer current.
The reduction procedure is described by Larsson et al. (this volume). 

\section{Results}

\begin{figure}[t]
  \resizebox{\hsize}{!}{
  \rotatebox{00}{\includegraphics{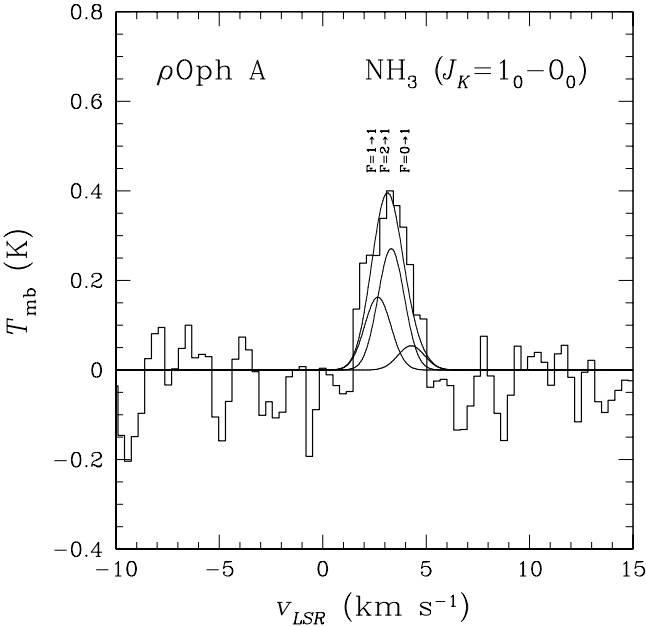}}
                        }
  \caption{The submm NH$_3$ ($J_{K}$\,=\,$1_0-0_0$) line (523.66\,\um) observed with Odin toward \roa\ (histogram).
  At 572.5\,GHz the Odin $T_{\rm mb}$-scale
  is related to the flux density by $F_{\nu}/T_{\rm mb} = 2600$\,Jy/K. 
  The quadrupole hyperfine lines in their equilibrium ratios are indicated, together with their total contribution.}
  \label{NH3_line}
\end{figure}

The $T_{\rm mb}=0.4$\,K line ($\eta_{\rm mb} = 0.9$; Hjalmarson et al., this volume) of NH$_3$\,($1_0 - 0_0$) toward \roa\
is centered on \vlsr\,$=3.2 \pm 0.1$\,\kms. The width of the hyperfine components (\cite{townes1955}) in Fig.\,\ref{NH3_line}
is 1.5\,\kms\ and some line broadening results from velocity smearing. 
In order to discuss the implications of this Odin observation we will first derive the physical characteristics of the source from 
ground based observations, specifically obtained for the Odin mission. 
The technical details of these SEST observations will be presented elsewhere.

\begin{table*}
\begin{flushleft}
 \caption{\label{sest_mol} A- and E-state methanol (CH$_3$OH) observations toward \roa\ with the 15\,m SEST. $\eta_{\rm mb}$ and beam-FWHM are, respectively, 
 0.73 and 52\asec\ for the $(2-1)$ and 0.67 and 35\asec\ for the $(3-2)$ transitions.}
\resizebox{\hsize}{!}{   
 \begin{tabular}{lccl ccc c lll} 
  \hline                          &                      		& 		 	&					& \multicolumn{3}{c}{Observation$^a$}& & \multicolumn{3}{c}{CH$_3$OH Model}\\
\cline{5-7} \cline{9-11} 
      Rotational                  &  Frequency$^{b}$     		&  $E_{\rm u}/k$	& \powten{6}$A_{\rm ul}$ 		& \vlsr     	  	& $\Delta v$ 		& $T_{\rm mb}$ 	 	& & $T_{\rm R}$ & $\tau_0$ 	& $T_{\rm ex}$	\\
      transition                  &  (MHz)               		&  (K)	         	& (s$^{-1}$)				& (\kms)  	& (\kms)     		& (K)         	 	& & (K)         	& & (K) 					\\
  \noalign{\smallskip}
  \hline
  \noalign{\smallskip}

$2_{-1}-1_{-1}\,\,\,\,{\rm E}$     & \phantom{1}96739.39  		& 12			&\phantom{1}2.48 			&  		  	& $1.51 \pm 0.05$ 	& $0.9 \pm 0.1$   	& & 1.16  	& $-0.07$ 	& $-11.6$	\\
$2_0-1_0\,\,\,\,{\rm A}$           & \phantom{1}96741.42  		&\phantom{1}7		&\phantom{1}2.38			&  		  	& $1.52 \pm 0.04$ 	& $1.3 \pm 0.1$   	& & 1.14  	& 0.09    	& 16.3 		\\ 
$2_0-1_0\,\,\,\,{\rm E}  $         & \phantom{1}96744.58$^{\star}$	& 20			&\phantom{1}3.30			& $3.36 \pm 0.09$ 	& $1.54 \pm 0.20$ 	& $0.3 \pm 0.1$  	& & 0.21  	& 0.05  	& \phantom{1}7.8\\  
$2_1-1_1\,\,\,\,{\rm E}  $         & \phantom{1}96755.51		& 28 			&\phantom{1}2.48			&  		  	& $0.85 \pm 0.31$ 	& $\le 0.10 $ 		& & 0.04  	& 0.005   	& 11.6 		\\       	  
$3_1-2_1\,\,\,\,{\rm A}^+$         & 143865.79            		& 28			& 11.16					&  $\ldots$ 	  	&  $\ldots$ 		& $\sim 0.04^c$ 	& & 0.04  	& 0.005    	& 10.2 		\\
$3_0-2_0\,\,\,\,{\rm E}$           & 145093.75            		& 27			& 11.93					&  		  	& $1.32 \pm 0.14$ 	& $0.50 \pm 0.13$ 	& & 0.32  	& 0.04    	& 12.9 		\\   
$3_{-1}-2_{-1}\,\,\,\,{\rm E}$     & 145097.47$^{\star}$  		& 20			& 10.61					& $3.33 \pm 0.09$ 	& $1.53 \pm 0.09$ 	& $1.75 \pm 0.06$	& & 1.72  	& 0.12    	& 18.4 		\\   
$3_{0}-2_{0}\,\,\,\,{\rm A}$       & 145103.23           		& 14			& 12.23					&  		  	& $1.52 \pm 0.09$ 	& $2.15 \pm 0.15$ 	& & 2.28  	& 0.21    	& 15.9 		\\             
$3_{2}-2_{2}\,\,\,\,{\rm A}^-$     & 145124.41            		& 52			&\phantom{1}6.50			& 		  	&  			& $< 0.06$        	& & 0.002 	& 0.0003  	& \phantom{1}9.7\\           
$3_{2}-2_{2} \,\,\,\,{\rm E}$  (blend)& 145126.37         		& 36			&\phantom{1}6.63			&  		  	& $1.28 \pm 0.19$ 	& $0.25 \pm 0.06$ 	& & 0.071 	& 0.009   	& 11.6 		\\    
$3_{-2}-2_{-2}\,\,\,\,{\rm E}$ (blend)& 145126.37         		& 40			&\phantom{1}6.63			&  		  	&  			&                 	& & 0.007 	& 0.0004  	& 10.8 		\\         
  \noalign{\smallskip} 
  \hline
  \end{tabular}
    }
\end{flushleft}   
Notes to the table: \\
$^{a}$  Average of data for two positions, spaced by 30\asec\ north-south. \\
$^{b}$  Rest frequencies were adopted from \cite{lovas}. Tuning frequencies are identified with an asterisk. \\
$^{c}$  Low resolution spectrum (1.4\,MHz). This spectrum includes also DCO$^+\,(2-1)$ at a level of $\int \!T_{\rm mb}\,{\rm d}v_{z} = 6.4$\,K\,\kms.
\end{table*}

\section{Discussion and conclusions}

\subsection{Physical parameters of \roa}

\subsubsection{\molh\ column density from C$^{17}$O\,($J = 1-0$)}

The SEST observations of \roa\ in the C$^{17}$O\,(1-0) line (46\asec\ beam, $\eta_{\rm mb}=0.70$) revealed
spectra with partially resolved hyperfine components. Their relative intensities reflect the ratios
of their statistical weights (0.5, 1.0, 0.75), indicating that the levels are populated according to their 
thermodynamic equilibrium values and that the emission is most likely optically thin. The lines are relatively 
narrow, \dv\,$=(1.29 \pm 0.05)$\,\kms, with the main line centered on \vlsr$\,=(3.41 \pm 0.02)$\,\kms. 
From these observations, the beam averaged column density of molecular hydrogen can be estimated from the standard
solution of the `radio'-equation of radiative transfer, i.e.
$N({\rm H_2}) = 2.75 \times 10^{12}\,\Phi (T_{\rm k}) \int \!T_{\rm mb}\,{\rm d}v_{z}\,\,\,({\rm cm}^{-2})$, where
$\Phi (T_{\rm k}) = 8 \pi k^3 T^2_{10} \exp{ (T_{10} /  T_{\rm k})} Q(T_{\rm k})/h^3 c^3 g_1 A_{10} \left [ 1 - J_{\nu}(T_{\rm bg})/J_{\nu}(T_{\rm k})\right ]$.
$T_{10} = h \nu_{10}/k = 5.390\,{\rm K}$ is the `transition temperature', 
$Q(T_{\rm k}) \sim  k\,T_{\rm k} / h B$ with $B = 56.1830$\,GHz is the partition function,
$g_1 = 3$ is the total statistical weight of the upper level $J=1$, 
$A_{10} = 7.13 \times 10^{-8}$\,s$^{-1}$ (\cite{chandra96}) is the spontaneous transition probability,
$T_{\rm bg} = 2.74$\,K is the temperature of the background radiation field and
$J_{\nu}(T) = T_{10} / \left [ \exp(T_{10}/T) - 1 \right ]$. 
The numerical constant, $2.75 \times 10^{12}$\,s\,cm$^{-3}$\,K$^{-1}$,
assumes the relative abundance of C$^{17}$O (with respect to \molh), $X({\rm C^{17}O}) = 3.6 \times 10^{-8}$, 
which is consistent with C$^{18}$O/C$^{17}$O\,=\,4 in \roa\ determined by \cite{encrenaz73}
(see also \cite{bensch01} for \ro\,C). 

The \molh\ column density is not very sensitive to the assumed temperature: 
the function $\Phi (T_{\rm k})$ varies within a factor of less than three (2.65) 
for $T_{\rm k}$ in the range 5\,K to 50\,K. For the observed line intensity 
$\int \!T_{\rm mb}\,{\rm d}v_{z} = (1.85 \pm 0.06)$\,K\,\kms, the \molh\ column density
is then in the range $(0.5 - 1.3) \times 10^{23}$\,\cmtwo. On the arcminute scale, this translates to
an average volume density of the order of $n({\rm H}_2) = (0.4 - 1.0) \times 10^{6} $\,\cmthree. 
These results are in accord with earlier molecular line observations (e.g. \cite{loren90}). 

\subsubsection{Kinetic gas temperature from CH$_{3}$OH ($J = 2-1$) and ($J = 3-2$)}

The observed spectra of 11 CH$_3$OH lines (Table\,\ref{sest_mol}; for an energy level diagram, 
see \cite{nagai79}) are suggestive of gas of relatively low
excitation (the ${\rm A}^-\,(3_{2}-2_{2})$ line with $E_{\rm u}/k = 52$\,K is not detected
and the ${\rm E}\,(3_{2}-2_{2}),\,(3_{-2}-2_{-2})$ blend, having $E_{\rm u}/k \ge 36$\,K, is weak, if
detected at all). We use large velocity gradient models (LVG) of methanol to obtain estimates of the average 
conditions in the core by requiring acceptable models to be consistent with the C$^{17}$O observations. 

We consider the rotational $J_k$ energy levels for both A- and E-type methanol in their ground torsional states. 
The level energies and  frequencies were obtained from the JPL-catalogue (\cite{pickett98}). 
For the A-states we adopted the Einstein-$A$ values from \cite{pei88}, whereas we calculated 
those of the E-states using the equations of \cite{lees73}, with appropriate H\"onl-London factors 
for the $a$- and $b$-type transitions (e.g. \cite{zare86}). Rate coefficients for collisions with He were 
kindly provided by D.\,Flower (see: Pottage et al. 2001, 2002) and were scaled for collisions with \molh\ 
($\times 1.37$).

The observed line spectrum (Table\,\ref{sest_mol}) is consistent with a model having $T_{\rm k} = 20$\,K, 
$N({\rm H}_2) = 6.0 \times 10^{22}$\,\cmtwo, $n({\rm H}_2) = 4.5 \times 10^5$\,\cmthree,
${\rm d}v/{\rm d}r = 45$\,\kms\,pc$^{-1}$ (1.4\,\kms\ along 40\asec), and a methanol abundance of 
$X($CH$_3$OH$) = X_{\rm A} + X_{\rm E} = (1.4 + 1.3)\,\times 10^{-9}$. The model yields thus a 
ratio $X_{\rm E}/X_{\rm A} = 0.89$, which can be compared to the equilibrium ratio of 0.67 at 20\,K.

Any beam effects of significance are not evidenced by the CH$_3$OH data (35\asec\ and 52\asec\ beam sizes), 
suggestive of emission regions not exceeding half an arcminute. Even the strongest methanol lines 
have only modest optical depths ($\max {\tau_0} \le 0.2$) and the excitation of these lines is only mildly subthermal, 
giving confidence to our temperature determination. Much higher temperatures (say 50\,K) are not consistent with 
the methanol observations. $T_{\rm k}$ is also equal to the temperature of the {\it cold} 
dust component evidenced by ISO-LWS observations (\cite{liseau99}).

\subsection{NH$_3$ ($J = 1-0$) emission from \roa}

The model of the previous sections represents the basis for our analysis of the Odin ammonia line observations,
where we varied only the NH$_3$ abundance. Using
the equations of \cite{poynterandkakar75} (and their `15 parameter exponential fit') we computed the level energies 
and Einstein-$A$ values, with the dipole moment from \cite{cohenandpoynter74}. 
The collisionial rate coefficients were adopted from \cite{danby88}. 

The observation with Odin (Table\,\ref{nh3_model}) can be fit with an ortho-ammonia abundance
of, formally, $X_{\rm o}({\rm NH}_3) =  4.25 \times 10^{-10}$, corresponding to a beam averaged column density 
$N_{\rm o}({\rm NH}_3) =  2.6 \times 10^{13}$\,\cmtwo. \roa\ has been mapped in the ($1_1$-$1_1$) and ($2_2$-$2_2$) 
inversion lines of NH$_3$ with the 100\,m Effelsberg
antenna (43\asec\ beam) by \cite{zeng84}. Their Fig.\,2 displays the spectra toward one position, and 
the values scaled to the Odin beam size
are given in our Table\,\ref{nh3_model}, together with our model for an ortho-to-para ratio of unity.
According to \cite{zeng84}, extended ammonia emission on the 90\asec\ scale is also observed. 
Fortuitously perhaps, a 43\asec\ source (\about\,45\asec\ SE)
is also consistent with the ($2_2$-$2_2$) and ($3_3$-$3_3$) observations by \cite{wootten94} with the VLA (6\asec\ beam). The
observed and model values are, respectively, 11.5\,K and 11.8\,K for ($2_2$-$2_2$) and 2.2\,K and 2.0\,K for
($3_3$-$3_3$), which is slightly masing. No data are given for their ($1_1$-$1_1$) observations, the model value of which is 45.2\,K. 
However, \cite{wootten94} used their ($1_1$-$1_1$) measurement (in combination with $2_2$-$2_2$) to estimate 
$X_{\rm p}({\rm NH}_3) \sim 3 \times 10^{-10}$. Albeit referring to a much smaller angular scale, this is in 
reasonable agreement with our ad hoc assumption of equal amounts of ortho- and para-NH$_3$. 

We thus estimate a total ammonia abundance of the order of 
$X({\rm NH}_3) =  8.5 \times 10^{-10}$ in the \roa\ core. This value is significantly lower than the 
\powten{-8} to \powten{-7}
generally quoted for molecular clouds (and cannot be explained by an erroneous ortho-to-para ratio) and 
contrasts with our value of $X$(CH$_3$OH), which appears entirely `normal' (e.g., \cite{ewine93}).
Although the uniqueness of the present model may be debatable (e.g., gradients in velocity, temperature and density
are known to exist over the 2\amin\ Odin beam), these results are certainly significant. 
Smaller velocity gradients, higher temperatures and/or densities would result in even lower $X$(NH$_3$).
At the other extreme, `normal' abundances would require the line to be thermalized ($\tau > 10^3$), e.g., 
at the unrealistically low value of $T_{\rm k} = 6$\,K, unless the NH$_3$ source is point-like. With this caveat in mind,
we conclude that the \roa\ core likely has a very low NH$_3$ abundance.


\begin{table}
\begin{flushleft}
 \caption{\label{nh3_model} Model results for NH$_3$ observations toward \roa}
\resizebox{\hsize}{!}{   
\begin{tabular}{lc lcl cl} 
  \hline
            &              & \multicolumn{3}{c}{NH$_3$ Model$^a$} &     &   \\
\cline{3-5}  
Transition  & $T_{\rm mb}$ &  $T_{\rm R}$ & $\tau_0$ & $T_{\rm ex}$&  $A_{\rm ul}$  & $n_{\rm c}$(\molh)$^{b}$  \\
            & (K)          &  (K)         &   & (K)         &  (s$^{-1}$)    & (\cmthree)               \\
  \noalign{\smallskip}
  \hline
  \noalign{\smallskip}      
o-$(1_0-0_0)$      & 0.40    &   0.41 &  4.6     & \phantom{1}6.5   &  $1.59 \times 10^{-3}$  & $3.6 \times 10^{7}$ \\
p-$(1_1-1_1)^{c}$  & $>0.38$ &   0.88 &  0.10    & 11.8             &  $1.69 \times 10^{-7}$  & $2.0 \times 10^{3}$ \\ 
p-$(2_2-2_2)^{c}$  & $>0.18$ &   0.23 &  0.012   & 21.7             &  $2.26 \times 10^{-7}$  & $2.0 \times 10^{3}$ \\  
o-$(3_3-3_3)$      & $\ldots$&   0.04 & $-0.005$ & $-5.2$           &  $2.59 \times 10^{-7}$  & $2.2 \times 10^{3}$ \\         

  \noalign{\smallskip} 
  \hline
  \end{tabular}
  }
\end{flushleft}   
Notes to the table: \\
$^{a}$ Ortho-to-para of unity is assumed ($X_{\rm o}/X_{\rm p} = 1$). \\
$^{b}$ Values of the critical density, $n_{\rm c}$(\molh)\,=\,$A_{\rm ul}/C_{\rm ul}(T_{\rm k})$, for 20\,K. \\
$^{c}$ Data estimated from Fig.\,2 of \cite{zeng84} and scaled to the Odin beam size, i.e. multiplied by $(43/120)^2$. 
\end{table}


\end{document}